# Application of a Coupled Harmonic Oscillator Model to Solar Activity and El Niño Phenomena

**Yasushi Muraki**

Institute for Space-Earth Environment Research, Nagoya University, Nagoya 464-8601, Japan

Abstract:   Solar activity has an important impact not only on the intensity of cosmic rays but also on the environment of Earth.   In the present paper, a coupled oscillator model is proposed in order to explain solar activity.   Using this model, the 89-year Gleissberg cycle can be naturally reproduced.   Furthermore, as an application of the coupled oscillator model, we attempt to apply the proposed model to the El Niño-Southern Oscillation (ENSO).   As a result, the 22-year oscillation of the Pacific Ocean is naturally explained.   Finally, we search for a possible explanation for the coupled oscillators in actual solar activity.

Keywords: 11-year solar cycle, Gleissberg cycle, ENSO, PDO, coupled oscillator model

1. Introduction: Motivation of the present research

   The present study was conducted considering the following background.   An 11-year periodicity and a 22-year periodicity have been found in the growth rate of tree rings in trees that have survived on Yaku Island for 1,924 years (Muraki et al. 2015).   Yaku Island is located at 30°E20'N 130°30'E in southern Kyusyu, Japan.   Quite surprisingly, the 11-year and 22-year periodicities are found during the Wolf Minimum (1290-1350 CE), the Spörer Minimum (1440-1550 CE), the Maunder Minimum (1650-1725 CE), and the Dalton Minimum (1810-1840 CE).   These periods are well known as grand minima of solar activity.

   In order to investigate why the 11-year and 22-year periodicities are found in tree rings, we analyzed meteorological data for Yaku Island.   Fortunately, records since 1938 exist for temperature, rainfall, and daylight hours.   These data are available from the Japan Meteorological Agency (JMA 2017).   When we analyzed the data, we could find no periodicity in the temperature or rainfall data, except for a one-year periodicity.   However, in the daylight hours (insolation) dataset, we have found an 11-year and a 22-year periodicity.

The 11-year periodicity was found in June when the island is covered by a thick cloud due to a monsoon, whereas the 22-year periodicity was observed in July and August when the island is usually covered by subtropical high atmospheric pressure.  Cloudiness affects the growth rate of cedar trees in terms of photosynthesis.  Therefore, we considered that the amount of clouds over the island may be affected by solar activity.  However, the origin of the 22-year periodicity remained unclear.

The remainder of the present paper is organized as follows.  In the next section, we explain the coupled harmonic oscillator model and present an application to solar activity.  The coupled oscillator model is then applied to the El Niño-Southern Oscillation (ENSO).  We demonstrate that the 22-year periodicity found in the Yaku cedar tree rings may arise from the ENSO.  Finally, we discuss the origin of the coupled harmonic oscillator model by comparing recent helioseismological observations.

## 2. Application of the coupled oscillator model to solar activity

First, we assume that there are two oscillators with slightly different periodicities in the Sun. The solar magnetic activity can be described by a superposition of the amplitude of each oscillator.

Let us refer to the two oscillators having slight different angular frequencies of $\omega_A$ and $\omega_B$ as Oscillator-A and Oscillator-B, respectively.  Then, the respective oscillations may be described by $\sin(\omega_A \cdot t)$ and $\sin(\omega_B \cdot t)$ waves.  Taking into account each phase ($\alpha$ and $\beta$), these oscillators may be expressed as $\sin(\omega_A \cdot t + \alpha)$ and $\sin(\omega_B \cdot t + \beta)$, respectively.  For simplicity, we take $\beta = 0$, and the absolute amplitude of each oscillator is nearly the same and is set to 1.0.  Then, the combined amplitude of the two oscillations is given by

$$\psi = \sin\{(\omega_A \cdot t + \alpha) + \sin(\omega_B \cdot t)\} = 2\sin\{(\omega_A + \omega_B)/2 \times t + \alpha/2\} \cdot \cos\{((\omega_A - \omega_B)/2 \times t + \alpha/2\}.$$

Herein, we let $T_A$ and $T_B$ denote the longest solar cycle and the shortest solar cycle observed between 1700 and 2015 CE ($\omega_A = 2\pi/T_A$, $\omega_B = 2\pi/T_B$), which are 12.05 years and 9.52 years, respectively.  These values were selected based on a Fourier analysis of the 315-year sunspot activity.  The results of the Fourier analysis are shown in Fig. 1.  We selected the above values from the crossing points of the $2\sigma$ line (95% confidence level).  The corresponding angular frequencies $\omega_A$ and $\omega_B$ are 0.521 and 0.660, respectively.

Here, the term $\sin\{(\omega_A + \omega_B)/2 \cdot t + \alpha/2\}$ expresses an average 10.6-year periodicity, whereas the term $\cos\{((\omega_A - \omega_B)/2) \cdot t + \alpha/2\}$ corresponds to a 89-year periodicity. The 89-year periodicity is known as the Gleissberg periodicity (Peristykh and Damon 2003). Interestingly, this periodicity has been independently detected in another sample of Yaku cedar tree rings based on the C13 measurements (Kitagawa and Matsumoto 1995). However, we cannot reproduce the 22-year periodicity. In Fig. 2, we compare our simple expression with actual sunspot data.

## 3. Application of the Coupled Oscillator Model to the ENSO

The ENSO is observed as an oscillation of the equatorial ocean temperature between the east coast of Indonesia and the west coast of Peru. We examine the possibility that the ENSO could produce the 22-year periodicity by the coupled oscillator model. Details on the ENSO may be found elsewhere (Hare and Mantua 2000, Barlow et al. 2001, Schneider and Miller 2001). A 5.35-year periodicity and a 3.55-year periodicity are chosen as the two fundamental oscillation frequencies (Guilyardi 2006). These numbers are obtained from actual observation data. Taking into account the spread of the actual data, the two periodicities are (5.35±0.6)-year and (3.55±0.25) years, respectively.

The combined amplitude of the two oscillations can then be expressed numerically as follows:

$$\psi = \sin\{(\omega_A \cdot t)\} + \sin(\omega_B \cdot t) = 2\sin\{(\omega_A + \omega_B)/2 \cdot t\} \cdot \cos\{((\omega_A - \omega_B)/2) \cdot t\}.$$

Here, the angular frequencies $\omega_A$ and $\omega_B$ are 1.769 and 1.185, respectively. From the term $\sin\{(\omega_A - \omega_B)/2 \cdot t\}$, we can obtain the 21.5-year periodicity. Taking into account the spread of the fundamental modes, the value is 21.5±5 years. Therefore, we believe that the 22-year periodicity found in the Yaku cedar tree rings may arise from ocean temperature oscillations rather than solar activity.

By the teleconnection process from the equator to the Pacific Ocean, several temperature oscillations may be induced over the ocean surface, one of which is the Pacific Decadal Oscillation (PDO) (Minobe et al. 2004). Another oscillation reported by Nitta is an ocean temperature oscillation arising from the La Niña phenomenon (Nitta 1990). When La Niña appears, the high tropical atmospheric pressure covering the Japanese coastal region will be suppressed. The frequent appearance of clouds is then expected over Yaku Island. This effect gives rise to variation in the growth rate of the tree rings via photosynthesis.

## 4. Meridional circulation of solar plasma and the coupled oscillator model

In this section, we discuss the correspondence between the two coupled harmonic oscillators and actual solar activity. Until now, we have extended discussions based on a complete phenomenological examination. Here, we compare the proposed model with actual observations of solar activity and attempt to understand which dynamics of the Sun may be related to the coupled oscillator model.

Zhao et al. (2013) reported a very interesting result on the solar convection zone based on the helioseismological data obtained by the Solar Dynamics Observatory (SDO)/Michelson Doppler Imager. According to Zhao et al., the meridional circulation of the plasma flow is formed by double layers. The plasma flow of the outer layer is carried from the equator poleward, and perhaps the flow returns back through the shallow interior path from the polar side toward the equator, the corresponding depth of which is r = 0.82 - 0.91$R_\odot$. At the deeper layer of the convection zone (r = 0.70 - 0.82$R_\odot$), the plasma is assumed to flow from the equator poleward (r = 0.70$R_\odot$) and probably takes the return path through the middle region of the convection zone (r = 0.85$R_\odot$). In the northern hemisphere, the plasma gas flows anticlockwise in the upper cell and clockwise in the lower cell. Up to this point, only observational results have been considered. Based on the above observation results, we hereinafter propose a model.

We assume that the currents in the upper cell and the inner cell constitute an oscillator and interact. In the northern hemisphere of the Sun, the toroidal current of the upper cell produces the poloidal magnetic field providing south polarity (S) around the North-pole region of the Sun. The toroidal currents of the outer and inner cells may constitute the coupled harmonic oscillator.

If we assume that the strength of the poloidal magnetic field at the north-pole region is on the order of $10^{-4}$ T, then the strength of the toroidal current can be estimated to be as large as $3\times10^{11}$ A by the following equation: $B = \mu I/(2\sqrt{8}\cdot R)$. Here, we adopt a value of $\mu = 4\pi\times10^{-7}$. Note, however, that as denser matter exits in the convection zone, $\mu$ requires correction. We set R to $6\times10^8$ m. On the other hand, in the parallel flow region at radius R = 0.85$R_\odot$, the distance between the two cells (plasma loops) may be rather short in comparison with the other region. When we assume that the distance between two cells is approaching 600 km, a much higher magnetic field of approximately 1,000 Gauss is expected to be induced by the turbulence, which may be on the same order as the magnetic field of the sunspots. The 22-year variation of the solar magnetic field may be induced by the competition of the two currents between the inner cell and the outer cell, and the overall feature of the solar magnetic field is determined. The coupled oscillator model is pictorially represented in Fig. 3.

Fig. 3 shows the solar minimum of 1986.  Here, the N- and S-magnetic poles, which correspond to the "plumbs" of the coupled harmonic oscillator, are pictorially represented.  In Fig. 3, the plumbs of the inner loop approach the closest distance, while the plumbs of the outside loop remain at the farthest distance.  At the equatorial region of the Sun, the embryos of the magnetic monopoles collide with each other and may cancel each other.  The energy of associated with this cancellation can be estimated to be $\sim 3\times 10^{22}$ Mx [gauss·cm$^2$].  A portion of the embryos will be transferred to the coronal mass ejection (CME).  The remaining portions will be pulled back by the attractive force between the N and S poles.

The correspondence described here may be compared with the relation between the thermodynamics and statistical mechanics of solar activity.  Thermodynamics describes thermal phenomena from a macroscopic viewpoint, whereas statistical mechanics explains them from a microscopic viewpoint.

The plasma circulating with the magnetic field may be an origin of the toroidal current. However, in the present study, we do not engage in the historical debate on the origin of the magnetic field (Babcock 1961, Hotta et al. 2015, Zharkova et al. 2015, Bercuz et al. 2015). Instead, we simply assume the presence of two toroidal currents in the inner and outer cells.  A schematic diagram is presented in Fig. 4 (Zhao et al. 2013).  In this model, a poloidal magnetic field with the opposite polarity (A$^-$) may be generated by the outer cell.  When the current in the outer cell becomes stronger, according to the law of Lentz, an inverse current may be produced in the inner cell.  Thus, the currents between the inner cell and the outer cell constitute a coupled harmonic oscillator.  There are two coupled oscillators in the Sun: one is located in the northern hemisphere, and the other in the southern hemisphere.

Fig. 5 presents a possible dynamo model for the solar cycles A$^-$ (image on the left-hand side) and A$^+$ (image on the right-hand side).  For example, when the solar poloidal magnetic field has negative polarity (A$^-$), the poloidal field is mainly induced by the toroidal current of the outer shell of the convection zone.  On the other hand, when the solar poloidal magnetic field has positive polarity (A$^+$), the global poloidal magnetic field may be induced by the internal current of the deep convection zone.  The strengths of these two currents are alternatively amplified by each other according to Lentz's law.  This may be the origin of the 22-year variation of the solar poloidal magnetic field.

Here, it would be interesting to compare the present model with other models, such as the dynamo model, on the magnetic field of Earth and the principle of a Tokamak nuclear fusion reactor. Although the construction of current Tokamaks is quite sophisticated, early studies considered an idea similar to the present model (Artsimovich 1967, 1968; Smirnov 2010; Chen 1974). The torus of a Tokamak may be compared with the ring current inside the convection zone of the present model. The Tokamak torus is surrounded by a poloidal magnetic field in order to confine the plasma within it. If an additional time-variable poloidal magnetic field is supplied, a ring current is induced inside the torus of the Tokamak by the induced electric field, *rot* **E**= − ∂**B**/ ∂t.

In the case of the Sun, the continuous energy to maintain the ring current may be supplied by the meridional flow of plasma. The plasma flows from the equator to the Polar Regions. According to actual observations, the meridional flow is very slow near the equator, whereas at higher latitudes its speed increases to approximately 10 m/s (Zhao et al. 2013). The magnetic flux is frozen-in the plasma. From the viewpoint of the plasma flow, the strength of the magnetic field is constant. In contrast, from a fixed point inside the banana-shaped cell of the meridional flow of the convection zone, the magnetic field varies with time, i.e., $\partial \mathbf{B}/\partial t \neq 0$. This may be a driving force for the toroidal current inside the convection zone (Zhao et al. 2014).

A similar result was obtained from recent calculations using a supercomputer for the origin of the magnetic field of the Earth. The result reveals the existence of possible alternative tubes of upstream and downstream flows in the outer core of the Earth (Kageyama and Sato 1997, Roberts and King 2013). A detailed discussion is not presented herein.

In closing, we consider two things. First, we consider the average speed of the meridional flow of the inner and outer loops. Taking the average speed of the inner flow as <v_in> ≈ 6.1 m/s and that of the outer flow as <v_out> ≈ 5.5 m/s, the 9.5-year and 12.1-year circulation periodicities of the plasma flow can be reproduced. Second, the sunspot embryos may be produced at r = 0.85⊙ at the middle latitude of the Sun when the two cells make their closest approach. The embryos of the magnetic poles (tiny magnets composed of plasma vortices) may be produced by the interaction of the two flows (turbulence). These embryos surface by magnetic levitation (Choudhuri et al. 1995, Charbonneau 2010) or by getting on the boiling plasma bubbles (Masada and Sano 2016).

5. Variation of cosmic ray intensity observed by neutron monitors and the proposed model

Here, we provide an interpretation of the cosmic ray intensity measured for a long period with the use of neutron monitors.  As known by cosmic ray physicists, the shapes of the intensity distributions of the solar minimum during the periods of 1969-1980 and 1992-2000, and those during the periods of 1960-1968, 1980-1991, and 2004-2012 are quite different, as shown in Fig. 6 (Hathway 2010).  The data obtained between 1969 and 1980 and between 1992 and 2000 have a "flat" distribution of cosmic ray intensity, whereas the data obtained during the other periods exhibit a sharp peak.  We interpret this difference based on the coupled oscillator model presented herein.  This difference may arise from the difference in the position of the oscillators, i.e., whether the solar minimum is induced by the outer oscillator or the inner oscillator.  The solar minimum is produced by the outer oscillator, as in 1986, and the magnetic poles are formed by the outside loop.  It would be interesting to determine whether the magnetic field produced by the outer oscillator rises and falls faster than that produced by the inner oscillator.

6. Summary

The conclusions of the present study are summarized as follows:

1. We have investigated a possible explanation for solar activity using a coupled oscillator model.
2. By taking the periods of the two oscillators as 12.1 years and 9.5 years, the 90-year Gleissberg cycle can be reproduced.
3. We speculate that the two oscillators represent the ring currents induced by plasma motion in the inner and outer loops of the meridional circulation.
4. Two types of variation of the cosmic ray intensity during the solar minimum may be related to the position of the oscillators, whether they are induced by the outer layer or the inner layer of the convection zone.
5. The coupled harmonic oscillator model can be applied to the ENSO.  The 22-year oscillation of the solar-environment can be reproduced as shown in Fig. 7.  The 22-year periodicity may be related to the ENSO, not the solar activity.  The coupled oscillator model has successfully explained other astrophysical objects (Osaki 1975, Aizenman et al. 1977).

Acknowledgements:

The author would like to thank Dr. T. Sekii of the National Astronomical Observatory of Japan (NAOJ), Profs. S. Shibata and H. Takamaru of Chubu University, Dr. H. Hasegawa of Kochi University, Prof. T. Shibata and Dr. S. Masuda of the Department of Environment of Nagoya University, and Prof. Emeritus of NAOJ, E. Hieii for valuable discussions.  The author would also like to thank Prof. M. Panasyuk of Moscow State University who kindly introduced us to Russian research on the solar dynamo mechanism presented at the 35th International Conference on Cosmic Rays at Busan, Korea in July 2017 and to Prof. Yu Yi for encouraging discussions.   The present study was supported in part by JSPS Grant-in-Aid (Kakenhi) No. 16K05377.

**References**

Aizenman A, Smeyers P and Weigert A, Avoided crossing of modes of non-radial stellar oscillation, Astr. Astrop. 58, 41-46 (1977).

Artsimovich IA et al., Sov. At. Energy 22, 325 (1967).

Artsimovich IA et al., 1968, Proc. 3rd Int. Conf. on Plasma Physics and Controlled Nuclear Fusion Researches (Novosibirsk, Russia, 1968) Nucl. Fusion (Special Suppl.) 17.

Babcock H, The topology of the Sun's magnetic field and the 22-year cycle, ApJ 133, 572-587 (1961)    DOI:10.1086/147060.

Barlow M, Nigam S and Berbery EH, ENSO, Pacific decadal variability, U.S. summer time precipitation, and stream flow, Journal of Climate 14, 2105-2128 (2001). http://dx.doi.org/10.1175/1520-0442 (2001)014<2105:EPDVAU>2.0.CO;2.

Berucz B, Dikipati M and Forgacs-Dajek E, A Babcock-Leighton solar dynamo model with multi-cellular meridional circulation in advection- and diffusion-dominated regimes, ApJ 806, 169 (18pp) (2015) doi:10.1088/0004-637X/806/2/169. (They predicts that the butterfly diagram has not yet been successfully reproduced by the dynamo calculation for double layers.)

Charbonneau P., Dynamo models of the solar cycle, Living Rev. Solar Phys. 7, 3-91 (2010) http://www.livingreviews.org/lrsp-2010-3.

Chen FF, *Introduction to Plasma Physics* (1974) (in Fig. 9.6 of Chapter 9), Plenum Press, ISBN 978-1-4757-0459-4.

Choudhuri AR, Schüssler M, and Dikpati M, The solar dynamo with meridional circulation, Astronomy Astrophysics 303, L29-31 (1995).

Guilyardi E, El Niño–mean state–seasonal cycle interactions in a multi-model ensemble,

J. Astron.Space Sci. (2018)    Application of coupled osccillator model to solar activity
Climate Dynamics 26, 329-348 (2006) DOI 10.1007/s00382-005-0084-6.

Hare SR, and Mantua NJ, Empirical evidence for North Pacific regime shifts in 1977 and 1989, Progress in Oceanology 47, 103-145 (2000).   https://doi.org/10.1016/S0079-6611(00)00033-1

Hathway DH, The solar cycle, Living Rev. Solar Phys. 7, 1-65 (2010). http://www.livingreviews.org/lrsp-2010-1 and arXiv:15020792v1 (astroph-SR) (Feb. 2015).

Hotta H, Rempel M, and Yokoyama T, High-resolution calculations of the solar global convection with the reduced speed of sound technique, The Astrophysical Journal 786, 24 (18pp) (2015) doi:10.1088/0004-637X/786/1/24.

Japan Meteorological Agency data base, (the English version of Yakushima metrological data can be downloaded from http://www.data.jma.go.jp/obd/stats/etrn/view/monthly_s3.php?prec_no=88&block_no=47836&year=&month=&day=&view=p4)

Kageyama A and Sato T, Generation mechanism of a dipole field by a magneto hydrodynamic dynamo, Physical Review E55,4617-4626 (1997).   DOI: https://doi.org/10.1103/PhysRevE.55.4617

Kitagawa H and Matsumoto E, Climate implications of $\delta^{13}C$ variations in a Japanese cedar (*Cryptomeria japonica*) during the last two millennia, Geophysical Research Letters 22, 2155-2158 (1995).   DOI 10.1029/95GL02066.

Masada Y and Sano T. Spontaneous formation of surface magnetic spectrum from large-scale dynamo in strongly stratified convection, ApJL 822, L22-L26 (2016). Doi:10.3847/2041-8205/822/2/L22

Minobe S, Schneider N, Deser C, Liu Z, Mantua N, Nakamura H, and Nonaka M, Pacific decadal variability: a review, submitted to Journal of Climate (2004) www.cgd.ucar.edu/staff/cdeser/docs/jclim_minobe-pdv.pdf.

Muraki Y, Mitsutani T, Shibata S, Kuramata S, Masuda K, and Nagaya K, Regional climate pattern during two millennia estimated from annual tree rings of Yaku cedar trees: a hint for solar variability? Earth, Planets and Space 67, 31 (6pp) (2015).   DOI 10.1186/s40623-015-0198-y.

Nitta T, Unusual summer weather over Japan in 1988 and its relationship to the tropics, Journal of the Meteorological Society of Japan 68, 575-587 (1990).   Also on the west coast of US side; see the paper of Horel JD and Wallace JM, Planetary-scale atmospheric phenomena associated with the southern oscillation, Mon. Wea. Rev. 109, 813-829 (1981).

Osaki Y, Nonradial oscillations of a 10 solar mass star in the main-sequence stage, Publ. Astrn. Soc. Japan 27, 237-258 (1975).

Peristykh AN, and Damon P, Persistence of the Gleissberg 88-year solar cycle over the last ~12,000 years: Evidence from cosmogenic isotopes, JGR 108, 1003 (15pp) (2003) doi:10.1029/2002JA009390.

9
-

Petrie GJD., Solar magnetic activity cycles, coronal potential field models and eruption rates, Astrophysical Journal, 768, 162-180 (2013), DOI: 10.1088/0004-637X/768/2/162

Roberts PH and King EM, On the genesis of the Earth's magnetism, Report on Progress in Physics, 76, 096810 (55p) (2013).   doi:10.1088/0034-4885/76/9/096801

Schneider N, and Miller AJ, Predicting western north Pacific ocean climate, Journal of Climate 14, 3997-4002 (2001) https://doi.org/10.1175/1520-0442(2001)014<3997:PWNPOC>2.0.CO;2.

Smironov NP, Tokamak foundation in USSR/Russia 1950-1990, Nuclear Fusion 50, 014003 (2010). doi:10.1088/0029-5515/50/1/014003

Zhao J, Bogart RS, Kosovichev AG, Duvall JR T., and Hartlep T, Detection of equatorward meridional flow and evidence of double-cell meridional circulation inside the Sun, The Astrophysical Journal Letters 774, L29 (6pp) (2013)   DOI:10.1088/2041-8205/774/2/L29.

Zhao J, Kosovichev AG, and Bogart RS, Solar meridional flow interior during the rising phase of cycle 24, Astrophys. J. Lett. 789, L7 (2014).   DOI:10.1088/2041-8205/789/1/L7.

Zharkova VV, Shepherd SJ, Popova E, and Zharkove SI, Heartbeat of the Sun from principal component analysis and prediction of solar activity on a millennium time scale, Sci. Rep. 5, 15689 (11pp) (2015)   doi: 10.1038/srep15689.

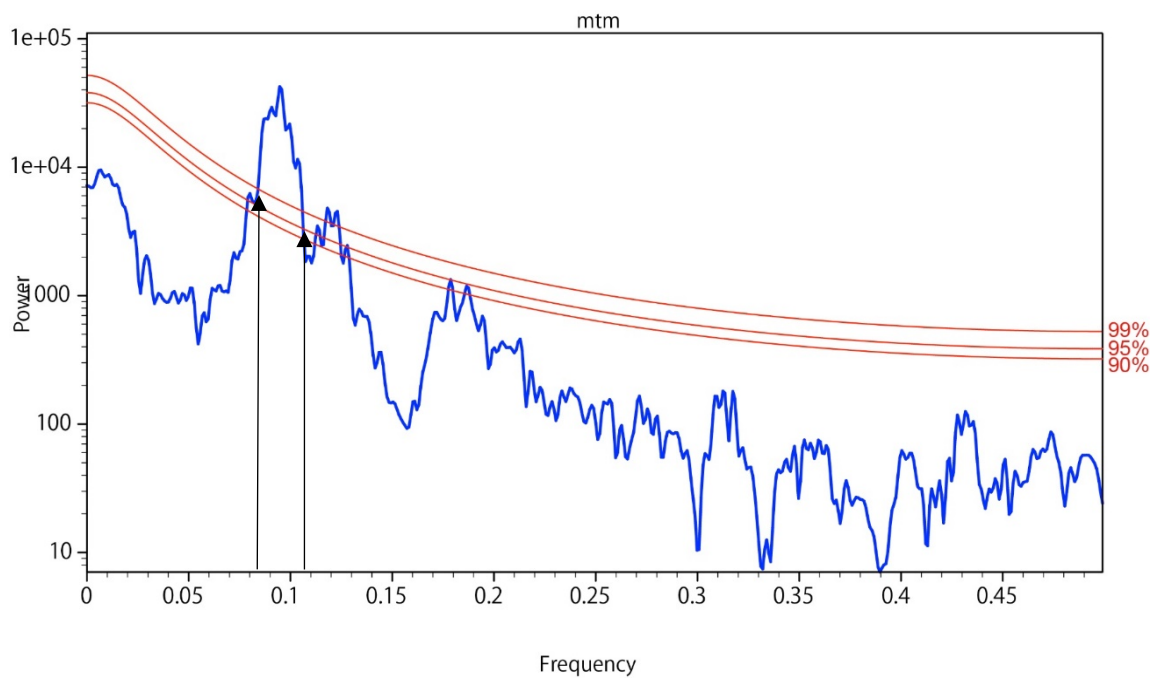

**Fig. 1**. Results of Fourier analysis for sunspots observed between 1700 and 2005. A sharp peak occurred far beyond the 99% confidence level, which corresponds to the 10.6-year periodicity. The arrows indicate periods of 12.1 years (left-hand side) and 9.5 years (right-hand side).

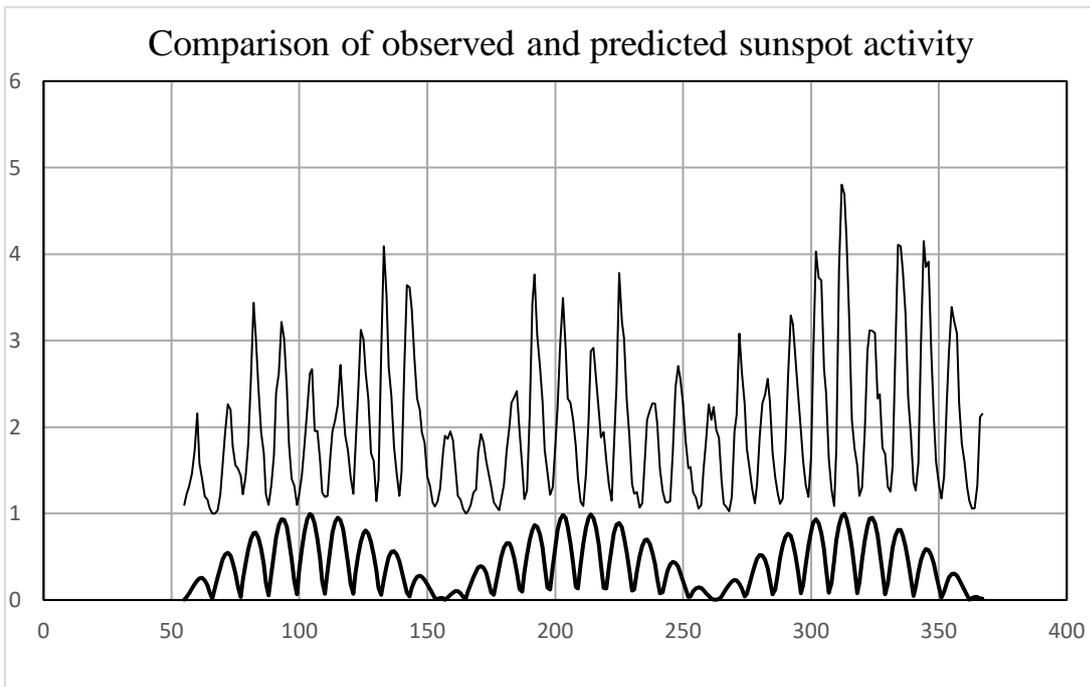

**Fig. 2.** Observed sunspot number (thin line) and sunspot number predicted by coupled harmonic oscillator model (thick line).   In order to adjust the start time, the prediction was shifted by 55 years.   The fundamental oscillator was produced with 22.0-year periodicity. However, for the purpose of comparison, the absolute value of the amplitude of ψ is taken. The maximum amplitude is normalized at 1.   Horizontal axis represents the year.

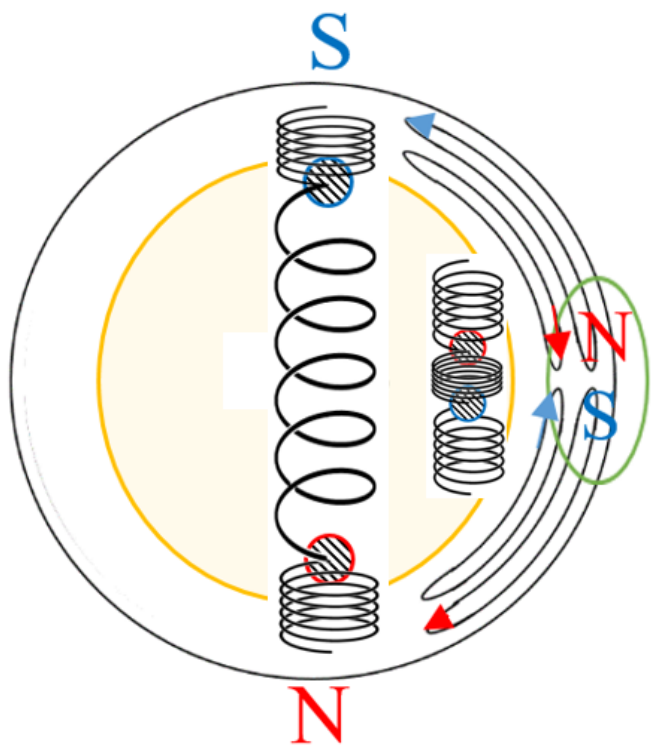

**Fig. 3**.  Schematic diagram of double coupled oscillator model.  The positions of the plumbs of the oscillator correspond to the magnetic poles of the solar magnetic field induced by the in*ner and outer* ring current*s*.  The image represents the solar minimum of 1986.  The inner oscillator (inner cell) approaches the minimum distance, whereas the outer oscillator (outer cell) are at the longest distance. The Gleissberg cycle is produced due to the slight different angular frequencies between the outer oscillator (Oscillator A) and the inner oscillator (Oscillator B).  The color of each pole is flowed by the notation of Petrie (2013).

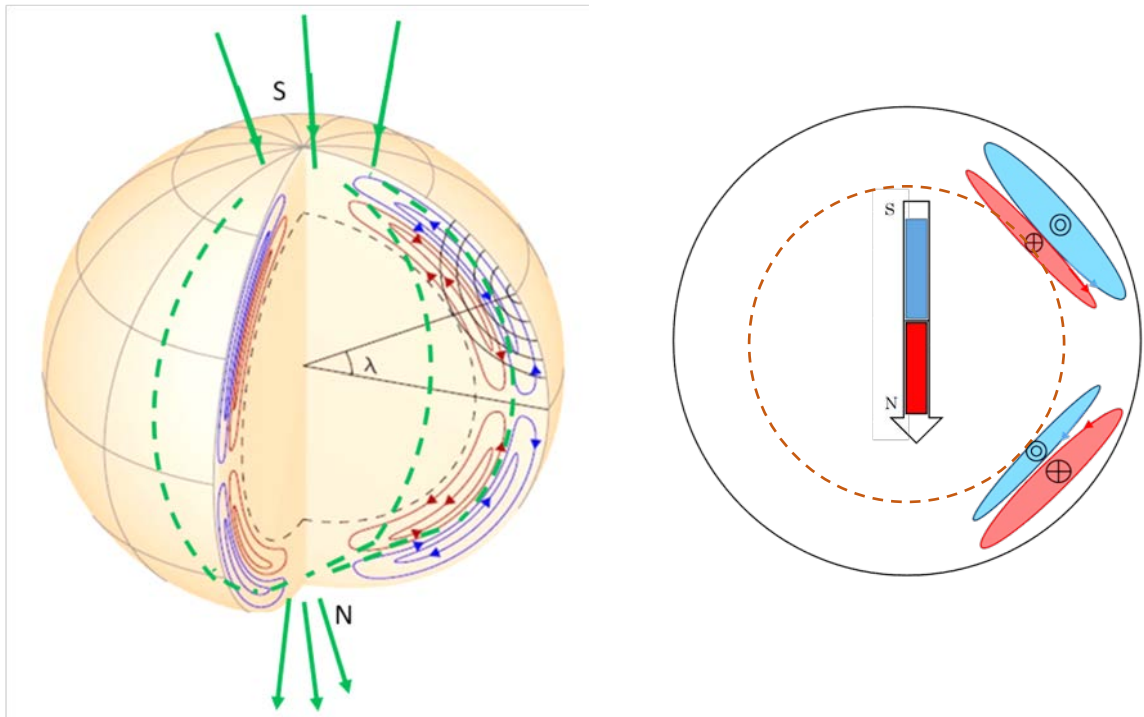

**Fig. 4**.   Pictorial representation of current model.   The left-hand side of the figure shows the poloidal field of the Sun induced by the outer cell, as shown in the figure of Zhao et al. (2013) (by permission of the AAS).   The green dotted lines show the poloidal magnetic field.   The image on the right-hand side shows the hypothetical toroidal currents inside the outer cell and the inner cell.   The blue shaded area with the double circle (◎) corresponds to the clockwise currents as viewed from the North Pole of the Sun, whereas the red ellipsoidal area with the cross circle (⊕) represents possible anticlockwise currents inside the convection zone.   The arrow in the central part of the picture indicates the direction of the poloidal magnetic field during solar cycles 22 and 24 (duration A$^-$ in Fig. 6), and the dashed circle corresponds to the border between the convection zone and the radiation zone of the Sun.

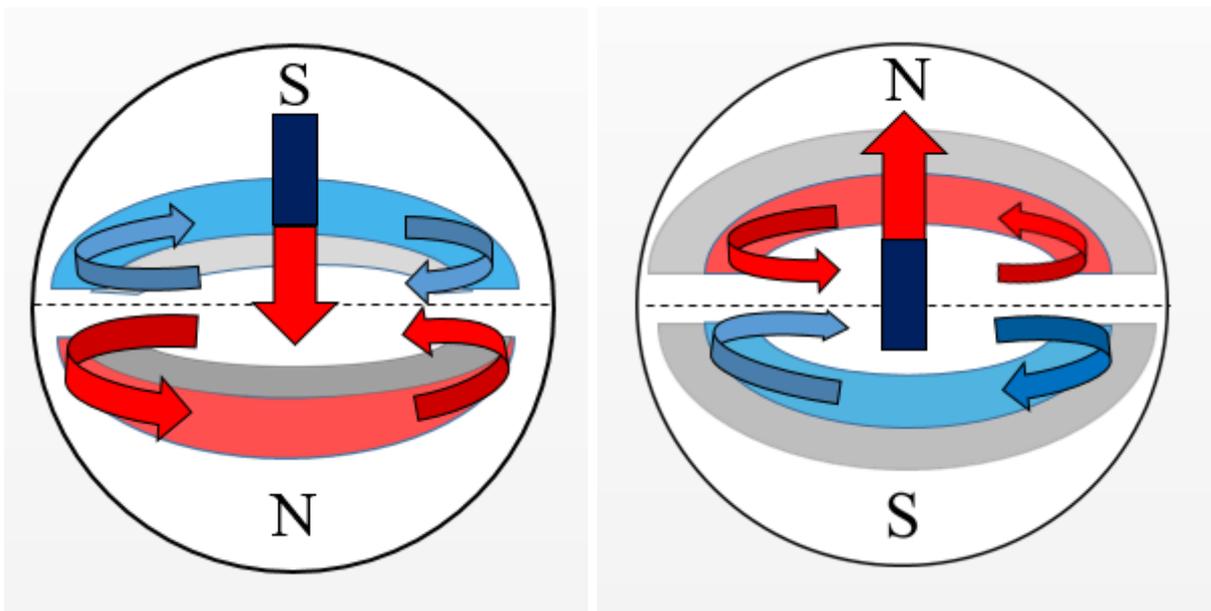

**Fig. 5**. Schematic diagram of solar magnetic field for negative polarity (A⁻) (image on left-hand side) and positive polarity (A⁺) (image on right-hand side).    For example, in the A⁻ case, the global poloidal current is induced by the clockwise current in the northern hemisphere and the anticlockwise current in the southern hemisphere as viewed from the North Pole. On the other hand, in the case of A⁺, the poloidal magnetic field is mainly induced by the internal currents in the convection zone.    A⁺ indicates the years 1954, 1976, and 1996, whereas A⁻ indicates the years 1964, 1986, and 2008, which include the solar minimum.

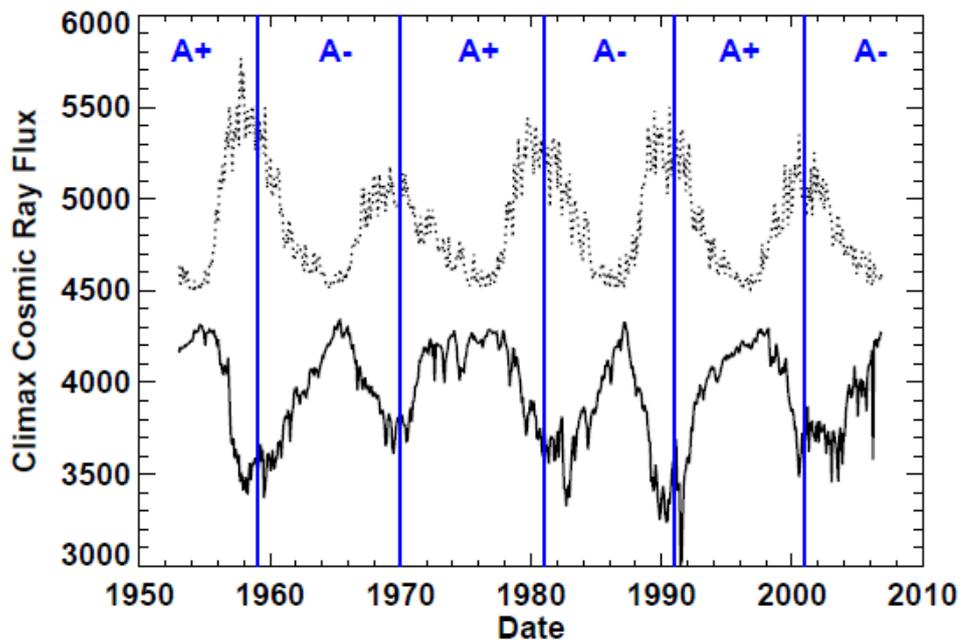

**Fig. 6**.   Dependence of cosmic ray intensity measured by Climax neutron monitor on sunspot number (dotted).    **A**+ indicates periods with positive polarity, and **A**- indicates periods with negative polarity.    This figure is courtesy of D.H. Hathaway, arXiv:15020792v1.

We suggest that the sharp rising and falling of the count rate during **A**- may be caused by the outer oscillator of the double combined harmonic oscillators.

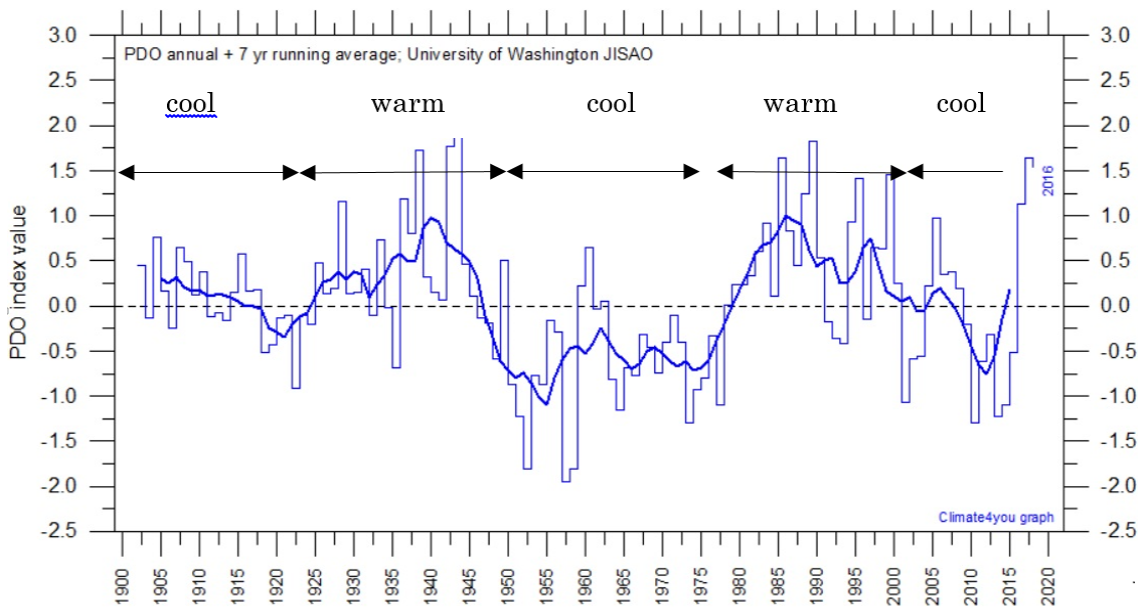

**Fig. 7**. Pacific Decadal Oscillation between 1900 and 2010.    The PDO phenomenon is unfamiliar to astrophysicists.    The surface temperature of the Pacific Ocean oscillates with an approximately 22-year periodicity between the high state (red) and the low state (blue) (difference of approximately ±2 degrees).    When the surface temperature of the California Coast is high, the state is defined as being positive (red).    This corresponds to the low state for the Japanese Pacific Coast.    For further details, see Minobe et al. (2004).    The horizontal arrows correspond to 22 years.